# Full-Sun observations for identifying the source of the slow solar wind

David H. Brooks[1,*], Ignacio Ugarte-Urra[1] & Harry P. Warren[2]

[1] College of Science, George Mason University, 4400 University Drive, Fairfax, Virginia 22030, USA

[2] Space Science Division, Naval Research Laboratory, 4555 Overlook Avenue SW, Washington, District Of Columbia 20375, USA

* Correspondence to: dhbrooks@ssd5.nrl.navy.mil

Present address: Hinode Team, ISAS/JAXA, 3-1-1 Yoshinodai, Chuo-ku, Sagamihara, Kanagawa 252-5210, Japan

## Abstract:

Fast (>700 km s$^{-1}$) and slow (~400 km s$^{-1}$) winds stream from the Sun, permeate the heliosphere and influence the near-Earth environment. While the fast wind is known to emanate primarily from polar coronal holes, the source of the slow wind remains unknown. Here we identify possible sites of origin using a slow solar wind source map of the entire Sun, which we construct from specially designed, full-disk observations from the Hinode satellite, and a magnetic field model. Our map provides a full-Sun observation that combines three key ingredients for identifying the sources: velocity, plasma composition and magnetic topology and shows them as solar wind composition plasma outflowing on open magnetic field lines. The area coverage of the identified sources is large enough that the sum of their mass contributions can explain a significant fraction of the mass loss rate of the solar wind.

## Introduction

Understanding the flow of energy and matter throughout the solar system is a fundamental goal of heliophysics, and identifying the solar sources of this flow would be a major step forward in achieving that objective. It would allow us to determine the physical properties of the plasma in the source regions, a significant constraint for theoretical models. Models of the solar wind, for example, are very sensitive to boundary conditions at the site of origin *(1)*.

The solar wind is comprised of a fast and a slow component *(2-4)*, both of which interact with and affect Earth's magnetic environment *(5)*. The origin of the fast wind is generally well established *(6,7)*, but there is still no consensus on the source of the slow wind. Many sources have been suggested, from helmet streamers *(8)* and 'blobs' disconnecting from their cusps *(9)*, to equatorial coronal holes *(10)*, active regions *(10,11)* and their boundaries *(12)* or chromospheric jets *(13,14)*. Narrow open-field corridors that connect coronal holes of the same polarity have also been proposed theoretically *(15)*. Recently, observations from the Hinode satellite have also identified specific outflow sites at the edges of active regions *(16-18)*.

Unfortunately, the two-component velocity structure of the solar wind cannot be used to distinguish sources low down in the solar atmosphere. The theoretically predicted *(3)*, and observationally confirmed *(2)*, high velocities of the wind at Earth are not observed in the low corona *(19,20)*, and the wind does not reach supersonic velocities until more than a solar radius above the Sun *(21)*, with acceleration not complete until at least 10 solar radii *(22)*. So, the acceleration to high velocities must take place at larger heights *(23)*. An important clue to the origin of the slow solar wind, however, is that the plasma composition (elemental abundance) is similar to that of the solar corona *(24)*, rather than to the solar photosphere, and this is a difference that can be exploited. The composition of the corona is enhanced with low first ionization potential (FIP) elements relative to the photosphere, and the degree of enhancement (fractionation) can be measured using the intensities of spectral lines from elements with different FIP, for example, Si (low FIP) and S (high FIP). Spectroscopic measurements of plasma composition from the Hinode EUV imaging spectrometer (EIS) have therefore become a valuable tool for attempting to establish a link between these candidate source regions on the Sun and *in situ* measurements at Earth, but this has only been achieved for the active region outflows and only for one active region observed in December 2007 *(refs 25, 26)*.

A significant problem with these measurements, however, is that the slow scanning time of spectrometers permits only limited field-of-view coverage, and so the presence of other, unobserved sources on the Sun at the same time cannot be ruled out. Estimates of the mass loss rate associated with the active region outflows, for example, suggest that they could account for 1/4 of the mass loss rate of the solar wind ($10^{12}$ g s$^{-1}$; ref. 16), but studies of other regions have shown that some portion of the mass may flow along large-scale closed loops and return to the surface in the vicinity of distant active regions (27), suggesting that the estimates may be too large. More recent measurements of velocities and densities in fact suggest that the mass loss rate may be overestimated by as much as an order of magnitude (25-28). So, even if individual outflow regions do contribute to the solar wind, the mass loss deficit needs to be made up from elsewhere on the Sun, implying that other sources are likely.

During 16–18 January 2013, we overcame this shortcoming using a new EIS-observing programme that scanned the entire Sun over a 48-h period, thus allowing us to map the whole disk to look for candidate sources. Despite being near solar maximum, the Sun was relatively quiescent during the scan, with only two flares that reached higher than Geostationary Operational Environmental Satellite (GOES) C-class according to the Hinode flare catalogue (29). One of these was associated with a (partial) halo coronal mass ejection (CME) that caused a temporary increase in the GOES proton flux, but both of the events occurred near the solar limb and were located far from where EIS was scanning at that time. The scan is based on a full-disk mosaic programme that we run every 3–4 weeks as part of efforts to monitor the instrument sensitivity in direct comparison with the Extreme Ultraviolet Variability Experiment on the Solar Dynamics Observatory (SDO). We execute a specially designed observing sequence at 26 positions on the solar disk by re-pointing the spacecraft to 15 positions and performing the scan using the top and bottom of the EIS CCD (charge-coupled device) as needed to cover the whole Sun. The programme takes a few hours to complete using the 40″ (arcsecond) slit. It is ordinarily not practical to run a similar sequence using the full spectral resolution EIS slits, however, because the scan would take several days and consume a large amount of telemetry.

For the purpose of this study, however, we designed a new EIS-observing sequence that matches the field of view of the regular scan (492″ by 512″) using the 2″ slit and coarse 4″ steps. The programme includes a series of Fe lines that we can use to measure the density and emission measure (EM—the distribution of plasma as a function of temperature). The specific Fe lines used are: Fe VIII 185.213 Å, Fe IX 188.497 Å, Fe X 184.536 Å, Fe XI 188.216 Å, Fe XI 188.299 Å, Fe XII 195.119 Å, Fe XII 203.72 Å, Fe XIII 202.044 Å, Fe XIII 203.826 Å, Fe XIV 264.787 Å, Fe XV 284.16 Å and Fe XVI 262.984 Å. The ratio of the Fe XIII lines at 202.044 Å and 203.826 Å is sensitive to electron density. The line list also includes Si X 258.37 Å and S X 264.22 Å that we can use to make abundance measurements. Their ratio is sensitive to the degree of fractionation of the plasma, when convolved with the EM distribution derived from the Fe VIII-XVI lines listed above. We and others have assessed the reliability of the line list extensively in studies of several different solar features *(30-33)*.

From our observations, we derive pure temperature images of the full Sun at the highest spatial resolution yet achieved, and Doppler velocity maps of the corona extending to higher temperatures than previously possible. We also compute the first plasma composition map of the entire Sun. By combining these observations with a magnetic field extrapolation model, we construct a unique slow-wind source map.

**Full-Sun images and plasma composition**
Figure 1 shows pure temperature images derived from spectral fits to the full-Sun data for a selection of the lines acquired by our observing programme. These images cover a range of temperatures from 450,000 K (0.45 MK) to 2,800,000 K (2.8 MK). Although similar images to Fig. 1 are routinely produced by the SDO Atmospheric Imaging Assembly (AIA), and instruments on other spacecraft, for a subset of these wavelengths, they are broad-band images with a spectral width of at least a few Å, which is significantly worse than the 0.0223 Å spectroscopic resolution of the images we show in Fig. 1, and thus contain multiple contributions from many different temperature spectral lines *(34)*. Spectrally pure temperature images with a broad temperature coverage are necessary to reduce the uncertainties associated with the integral inversion techniques that allow us to convert the line intensities into properties of the emitting plasma, such as electron density, EM or the degree to which the plasma composition is enhanced. We prepared coaligned full-Sun intensity images like that in Fig. 1 for all the spectral lines used in our analysis. Figure 2 shows an expanded image for the Fe XIII 202.044 Å line formed at 2 MK. We draw attention to this image because it is central to our analysis.

The ratio of the intensities of spectral lines from low FIP and high FIP elements can be used to calculate their relative abundances and thus the plasma composition for our full-Sun map. Previous studies *(25, 35)* have examined the abundance diagnostics in the EIS spectra and concluded that the Si X 258.37 Å/S X 264.22 Å ratio is one of the best. Compared with the other available ratios, it is relatively insensitive to the electron temperature and density. The variation is 30–40% in the temperature region where the lines are formed (around 1.4 MK). There is, however, a strong variation at high temperatures, and a factor of 2.3 sensitivity to density in the log $n$=8–10 range. We therefore need to measure the density to account for that sensitivity, and convolve the ratio with the EM distribution to account for any significant high-temperature emission. Following our previous work *(25)*, we performed these calculations for every pixel in the full-Sun data set by first deriving the electron density using the Fe XIII 202.04/203.83 ratio and then using that density to compute contribution functions (the equivalent of an imager filter's temperature response) for all the spectral lines in our observing programme. We used the CHIANTI database v.7 (*refs 36, 37*) assuming a photospheric composition for the plasma *(38)*. We then fit the observed intensities by convolving them with an EM distribution derived from a Monte Carlo simulation. The Monte Carlo code is available in the PINTofALE software package *(39, 40)* and uses a Markov-Chain algorithm to find the best-fit solution. Only the Fe lines were used for the EM calculation to minimize any uncertainties due to elemental abundances. Since Fe and Si are low FIP elements, their abundances (and hence intensities) are expected to be enhanced in the corona to a similar degree. Most of the Fe lines used for the EM analysis, however, lie on the short-wavelength (SW) EIS detector, whereas the Si X 258.37 Å and S X 264.22 Å lines lie on the long-wavelength detector. So, uncertainties in the cross-calibration of the detectors, and their evolution with time *(41, 42)*, could lead to a mismatch between the Fe EM and Si X 258.37 Å absolute intensity. We therefore scaled the derived EM distributions to ensure that the Si X 258.37 Å line is reproduced. This procedure also accounts for any uncertainties in the Fe/Si abundance. In our previous work *(25)*, we found that this scaling was always <20%, but that study used observations from early in the mission. So, we checked whether the method accounts for any sensitivity evolution in these more recent data by examining an area of one of the rasters where many of the pixels require larger scaling and re-calibrating the line intensities using two different methods that attempt to account for the sensitivity changes *(41, 42)*. Our experiment verified that the scaling was reduced to under 30%, which is comparable to the accuracy of the method (see Methods section). Using the derived distribution, we then calculated the degree to which the plasma is fractionated by computing the expected intensity of the S X 264.22 Å line. The ratio of the predicted to observed intensity for this line gives us our level of fractionation compared with photospheric values.

Given the size of the full-Sun data set, 16 million calculations were needed to produce a plasma composition map including every pixel over the full Sun. We show a display version of this map in Fig. 3, created from the ratio of Si X 258.37 Å and S X 264.22 Å lines. This image captures the main features of the composition map, such as whether a structure has photospheric or coronal abundances and the relative level of enhancement, but was not used for any of our analysis. We show this map for presentation because it is only affected by bad pixels or missing data in the Si X and S X lines, and this is relatively easier to filter out. The full composition map, however, is affected by bad pixels and missing data in all of the spectral lines from Fe VIII to Fe XVI and is therefore much noisier and more difficult to interpret visually. We stress that we used the full composition map created using all the spectral lines for all of the quantitative analysis.

We then filtered the full composition map to define areas with an enhanced (slow wind) composition. We used an enhancement threshold of 60% to include the entire range of fractionation values, which accounts for the fact that the slow wind has a variable composition *(24)*. It also attempts to account for the fact that the Si/S ratio does not always show a clear fractionation pattern. Although the EIS observations show that the Si/S ratio can detect variations in composition between, for example, polar coronal holes (photospheric composition) and active region outflows (coronal composition) *(25)*, the two elements lie close to the traditionally defined boundary between low and high FIP elements, and some models *(43)* suggest that the ratio may underestimate the enhancement factor due to possible underfractionation of Si and overfractionation of S. Here we only use the measurement to determine whether the plasma in a pixel is fractionated; the actual enhancement factor itself is not used in any computations. We stress, however, that the EIS composition measurements are in agreement with the general trends seen in the *in situ* data.

**Full-Sun velocity map**

Regions of enhanced composition could be possible sources of the slow speed wind, but our full-Sun composition map alone cannot show us whether the plasma from these regions is actually upflowing from the solar atmosphere. We obtained this information from the Doppler shift of the spectral line centroids, and derived radial velocity maps for several of the lines in our observations.

The EIS spectrometer has several peculiar characteristics that make Doppler velocity measurements difficult. For example, EIS does not observe any photospheric spectral lines, so it is not possible to obtain an absolutely calibrated wavelength scale. The Doppler velocities we use are therefore relative velocity measurements. Furthermore, the EIS slit is not perfectly aligned to the vertical axis of the CCD and there is a drift of the spectrum on the CCD due to thermal variations in the instrument around the satellite orbit. These effects have been extensively investigated and we accounted for them using our best current knowledge of the instrument *(28, 44)*. Most of the instrumental effects are corrected using the recommended neural network model *(44)*. This model uses the Fe XII 195.12 Å line as calibration standard and assumes that velocities in this line are 0 when averaged over the entire mission. There is some evidence that this assumption is not accurate *(45)* and that coronal lines may exhibit blue shifts of a few km s$^{-1}$. Therefore, some care needs to be taken when choosing a reference wavelength to calibrate the Doppler velocities. Here we refined the velocity measurements by correcting to an off-limb reference wavelength for the Fe XII 195.12 Å line *(28)*. This reference wavelength was obtained by averaging the line profiles measured in two large quiet regions above the East and West limbs where the spectral line is expected to be close to its rest (or slightly blue shifted) wavelength. The wavelength scale was then shifted to the reference wavelength, and the correction was then applied to the strong Fe XIII 202.04 Å line that is within the same wavelength band. The final velocity measurements have uncertainties of ~4.5 km s$^{-1}$, and they are converted to radial velocities using a simple cosθ expression. Here we only use radial Doppler velocities calculated from the Fe XIII 202.04 Å data.

We show an example velocity map in Fig. 4. It shows regions of plasma upflow and we can compare their locations with features in the intensity image in Fig. 2. We see, for example, that the bright active region in the North West hemisphere (AR11654) has large areas of upflow, but mostly on the solar Eastern side, suggesting that we can rule out the red-shifted downflow areas in and around the active region (AR) core as a solar wind source (in a direct sense).

**Open/closed magnetic field model**
The velocity and composition maps reveal the locations of slow-wind composition plasma that is upflowing, but the magnetic field topology is key to determining whether these upflows become outflows, really escape into interplanetary space and are directed towards the ecliptic plane where they can be measured *in situ*. No direct open magnetic field channel, for example, has been established for the December 2007 region *(46)*, and other cases where the magnetic topology has been inferred show that not all of the upflows can escape on open field *(47)*. Therefore, we also generated a full-Sun potential field source surface (PFSS) extrapolation *(48, 49)* to add this final piece to the puzzle.

The PFSS approximation has significant limitations. In particular, the corona is unlikely to be free of electric currents, and these alter both the strength and connectivity of the magnetic field compared with a potential configuration. Conversely, the model has been relatively successful in capturing the large-scale coronal field *(10, 49, 50)*, which is the objective here, and we only use it to determine whether a field line is open and whether it extends down towards the ecliptic plane. Our view is that the most likely shortcoming of our use of the PFSS model is that we may miss field lines that bend more dramatically towards the ecliptic. Definitive confirmation of our results will come from applying more sophisticated magnetic field models in the future.

We used the PFSS package available in SolarSoft *(49)*. This package allows access to a database of samples of potential field models (at 6 h cadence), constructed from Helioseismic and Magnetic Imager magnetogram observations, for any heliographic latitude and Carrington longitude. Field lines are then traced out from these locations until they either close back onto the Sun or open out to reach the source surface where they are forced to be radial. We extrapolated magnetic field lines from each of the EIS coordinates, corresponding to every pixel, using the nearest sample to the time of the centre of each EIS raster scan. This ensures that the extrapolation is always made from a magnetic field model sampled within 3.5 h (often much less), since the PFSS model generally does not evolve significantly during that time-frame. We then converted the EIS solar coordinates in arcseconds to heliographic coordinates, corrected for the solar B angle, and finally converted to Carrington angles.

We computed a total of ~1.6 million potential field lines from this model to cover the full Sun. We show a subset of these field lines in Fig. 5, overlaid on the intensity images in Fig. 1, and again in Fig. 6, overlaid on the velocity map in Fig. 4. Figure 5 clearly shows that many of the open-field lines are associated with active regions, and Fig. 6 shows, for example, that the Eastern outflow from AR11654 does indeed lie on open magnetic field lines, some of which extend down into the ecliptic plane.

From the PFSS extrapolation, we determined which magnetic field lines reach the source surface, and are therefore open, for every EIS pixel in the data set. These data can then be mapped to our Doppler velocity and plasma composition maps to find potential solar wind sources.

**Slow solar wind sources and mass loss rate**
We combined all this information to produce our solar wind source map, adopting a number of criteria to decide whether a pixel should be counted as a candidate slow-wind source and ultimately included in our mass loss rate calculation (see below). Some of these are purely technical. For example, we only included pixels if the numerical calculation of the EM distribution was well constrained. We set the condition that the $\chi^2$ should be no larger than the number of lines in the integral inversion. This ensures that the difference between the calculated and observed intensities is generally within the calibration uncertainty. Since the EM calculation depends on the density, we excluded values well outside the range of sensitivity of the ratio (8<log $n$<11). We also excluded pixels with a poor spectral fit to the Fe XIII 202.04 Å line.

A number of other criteria were also used. First, we only included pixels where the PFSS field line trace reaches the source surface, that is, the magnetic field is open. Second, as we show in Fig. 7, 90% of the mass flux comes from pixels below ~40°, so we only included pixels whose traced field line originated from below this latitude. Practically speaking, none of the field lines from above this latitude influence our study because they do not extend down close to the ecliptic plane, but it is unclear how close the other field lines must reach to be able to deliver mass flux to the ecliptic plane that can later be observed *in situ*. Most of the mass flux originates from above 11°, however (Fig. 7), which implies that some field lines from these latitudes should not be excluded. So we set this as the threshold, but it is clearly dependent on the model, which is of course simplistic. Third, we only included pixels within the solar radius on January 17 (midway through the scan). Pixels high above the limb do correspond to locations in the magnetic field models and a field line can be traced from them. But those locations rotate over the limb, not radially off-limb, so the back projection and radial velocity correction have increasing uncertainty close to and above the limb. Fourth, we assumed that the plasma was fractionated if the enhancement above photospheric levels was >60%; this is well above the radiometric calibration uncertainty (~23%; *ref. 51*).

The fractionation measurements also depend on updates to the radiometric calibration. Using an alternative re-calibration of the sensitivity evolution of the instrument since 2006 *(ref. 41)*, we calculated that <5% of the pixels would change by more than this amount. Finally, given the uncertainties in the radial Doppler velocities (4.5 km s$^{-1}$), we have to make a careful choice of velocity threshold to decide whether the plasma in a pixel is upflowing or not. There are two possible approaches: assume (1) that any motions along the line of sight average to 0 so that the coronal lines are at rest or (2) that they have a blue shift *(45)* of a few km s$^{-1}$. In case (1), the mass flux would be underestimated if they actually have a small blue shift because fewer pixels would be included, while case (2) would overestimate the mass flux if they are at rest because more pixels would be included and their velocities would also be larger. Both cases will underestimate the mass flux if we exclude pixels with velocities below 4.5 km s$^{-1}$. Fortunately, the identified source regions are not particularly sensitive to this choice: the main effect is that they become more extended and/or denser due to more pixels being included, but the mass flux calculation itself is significantly affected, as we discuss below.

We show the final map in Fig. 8. The red and green areas show regions where enhanced composition plasma is outflowing on open-field lines that extend down close to the ecliptic plane, and these areas meet all of the criteria that, in our view, make them possible sources of the slow-speed solar wind. Their area coverage is at least 50 times greater than the outflow area estimate by Sakao *et al. (16)*, which is more than enough to overturn the lower density and velocity measurements found recently for active region outflows when calculating the mass loss rate contribution to the slow wind. We calculated the total mass loss rate for our candidate sources using the formula

$$M = \sum_{i=1}^{N} m_p n_i v_i l^2$$

where $m_p$ is the proton mass, $n_i$ is the electron density of pixel $i$, $v_i$ is the radial velocity of pixel $i$, $l^2$ is the area of an EIS pixel and $N$ is the total number of pixels that meet all of our selection criteria. Using the Fe XIII densities and velocities, we measured the mass loss rate at every pixel, to include every possible type of source, and calculated the mass flux. As discussed, the mass flux depends on the choice of velocity threshold, which is illustrated in Fig. 9. In one extreme, we assume the coronal lines are at rest and only pixels with velocities above the uncertainty are included. In the other extreme, we assume the coronal lines have a small upflow of 1.5 km s$^{-1}$ and include all blue-shifted pixels. This leads to calculated mass loss rates of $1.5-2.5 \times 10^{11}$ g s$^{-1}$. Assuming an Earth-directed isotropic distribution, these measurements translate to proton flux densities at Earth of $6.6 \times 10^7 - 1.1 \times 10^8$ cm$^{-2}$ s$^{-1}$, which can be compared with the *in situ* measurements made in the days following our scan, when the plasma has had sufficient time to travel to Earth, by the Advanced Composition Explorer (ACE) Solar Wind Electron, Proton and Alpha Monitor (SWEPAM) *(52)*. The ACE data (Fig. 10) show that the radial solar wind velocity at Earth was ~400 km s$^{-1}$ during the 20–22 January period, which is typical of the slow wind. The proton flux density is quite stable, with an average of $1.3 \times 10^8$ cm$^{-2}$ s$^{-1}$, and 50–80% of this can be accounted for by the EIS measurements. The mass flux comparison and ACE velocity data also imply that there is no significant contribution from the fast solar wind during the observation period.

## Discussion

The comparison between Hinode and ACE observations obviously has large uncertainties because ACE makes measurements corresponding to the features that it actually sees along the Sun-Earth line and projected back to the surface, but since 90% of the outflow mass flux comes from below 40° latitude (Fig. 7), the comparison at least shows that the observed sources in the low corona can potentially supply enough mass flux into the heliosphere, and towards the ecliptic, to explain most of the actual *in situ* particle measurement, rather than a generic value for the solar wind mass loss rate.

We stress that other candidate sources in the low corona, such as magnetically confined plasma, un-fractionated photospheric plasma or downflowing plasma, cannot contribute directly to the wind: the plasma must first be fractionated to FIP bias levels measured *in situ* in the slow solar wind *(24)* and then expelled on open magnetic field lines. At that point, they would appear in a solar wind source map similar to ours, with exactly the same signature that we are showing. The remaining mass flux may be more likely to come from other sources that are only visible in the higher corona.

On the basis of our analysis, the majority of the mass flux from the low corona, however, appears to come from the edges of active regions (red areas in Fig. 8) *(10, 11, 17, 49, 53, 54)*. Like coronal hole boundaries, active regions can be bisected by the heliospheric current sheet *(46)*, so outflows from either side could explain why the heliospheric current sheet is always surrounded by slow wind *(55)*. A minority flux component also flows from a few coronal hole-like regions, whose shapes follow the boundary between quiet and active areas, but these sources are less concentrated. As such, they are less visible in our Figure, so we have highlighted them in green in Fig. 8. Our results support the view that the slow wind flows from several contributing sources: the red and green areas in Fig. 8 and an unknown source, possibly in the higher corona, that contributes the rest of the mass flux. The figure clearly shows many more red areas than green areas in the low corona, however, so active region outflows appear to be the primary source, at least for the time interval of our observations.

There are of course uncertainties in our results because of the methodology of the analysis. For example, we implicitly assume ionization equilibrium to perform the EM analysis, but this might be violated by the high-speed wind motion at the outflow sites. We do not know the driver of the outflows, but since the ionization relaxation time is less than ~200 s in the low corona *(56)*, equilibrium is reached fairly rapidly even if they are generated by some impulsive heating mechanism such as chromospheric jets or spicules. Departures from equilibrium could lead to observable effects, such as anomalously high intensities of Li-like spectral lines that have relatively longer relaxation timescales than spectral lines of other iso-electronic sequences *(57)*. As evidenced by the low values of $\chi^2$ in our calculations, we did not, however, detect any non-equilibrium-based discrepancy in our EM analysis. These effects would be interesting to investigate more systematically in the future by attempting to reproduce the observed intensities with coupled time-dependent ionization and hydrodynamic models of the outflows *(58)*. The longevity of the outflows (they can persist for several days *(25)*), however, and the fact that the FIP bias has had time to evolve to coronal values, possibly suggests that the outflows have reached a quasi-equilibrium state by the time they are detected in the corona, and are less likely to be the result of material being ejected rapidly from below the photosphere *(26)*.

In an ideal scenario, we would further constrain the sources by examining the *in situ* composition data from ACE/SWICS (Solar Wind Ion Composition Spectrometer), as we did in our previous study *(25)*. By filtering out all the regions in our map where the composition measured by EIS does not match that of SWICS, we could make a conclusive link between the wind sources on the Sun and the *in situ* plasma measurements. This comparison, however, was not possible for the time period of our observations, so we should remain cautious. A definitive analysis awaits future observations and will be a focus of the upcoming Solar Orbiter mission.

# Acknowledgements

D.H.B. thanks M.L. DeRosa for guidance on the use of the PFSS software. This work was performed under contract with the Naval Research Laboratory and was funded by the NASA Hinode program. Hinode is a Japanese mission developed and launched by ISAS/JAXA, with NAOJ as domestic partner and NASA and STFC (UK) as international partners. It is operated by these agencies in co-operation with ESA and NSC (Norway). CHIANTI is a collaborative project involving George Mason University, the University of Michigan (USA) and the University of Cambridge (UK).

**Figure 1: EIS images of the solar corona.**

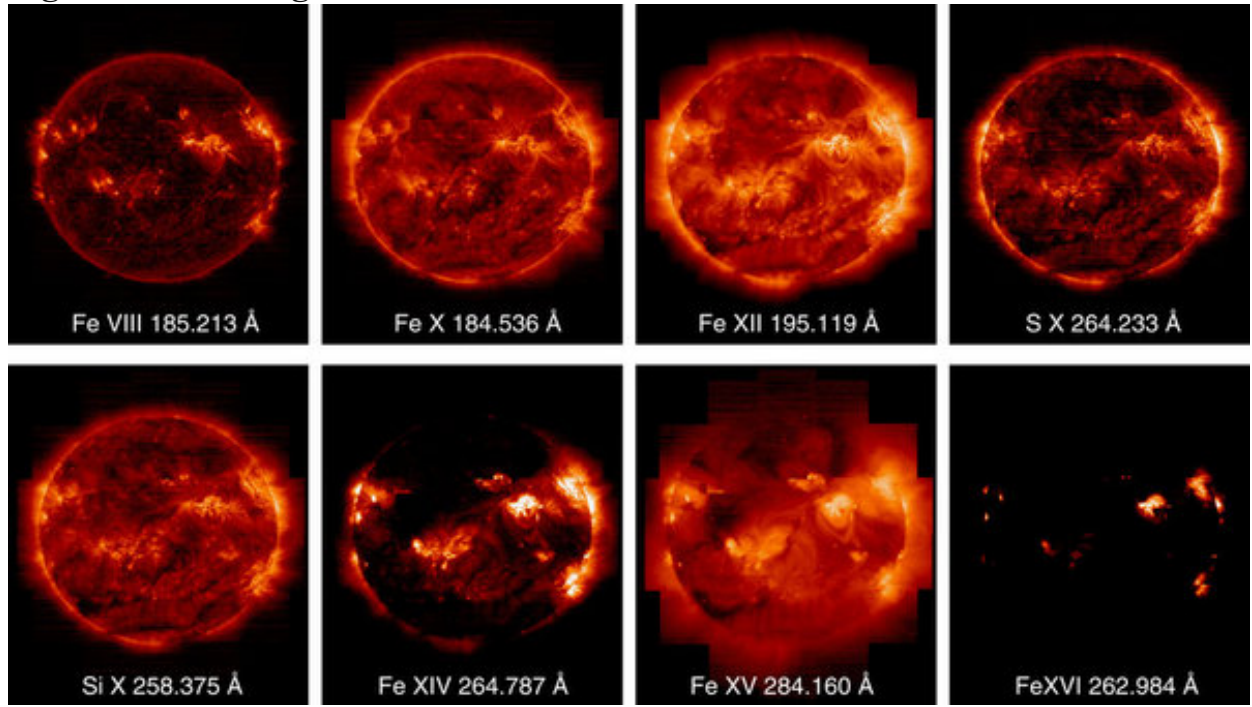

These pure temperature images are constructed from EIS spectral line intensities and cover a range of temperatures from 0.45 to 2.8 MK. They are used to construct EM distributions at every pixel. The mosaic is constructed by recording the top and bottom readings of the EIS CCD at 26 positions on the solar disk. The observing sequence scans across the central disk, around the North limb and finally around the South limb. It is sheared East-West in some locations because of solar rotation during the 2-day spectrometer scan. Details of the observing sequence are given in the main text.

**Figure 2: EIS image of the solar corona at 2 MK.**

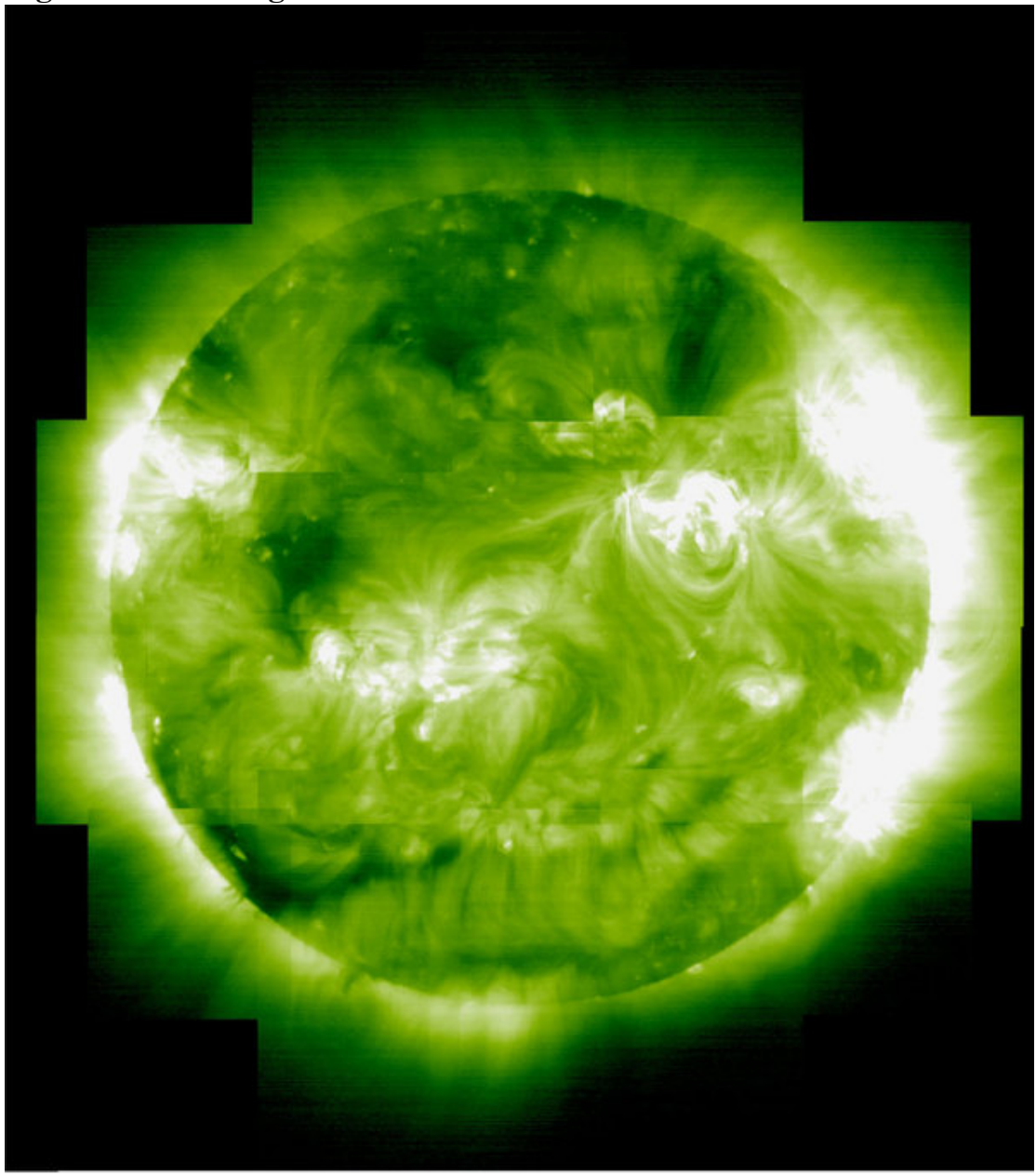

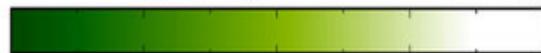

An expanded image of the solar corona at 2 MK constructed from Fe XIII 202.044 Å spectral line intensities. This line was used to determine Doppler velocities, and its ratio with Fe XIII 203.83 Å was used to compute electron densities. It is formed close in temperature to the Si X 258.37 Å/S X 264.22 Å ratio used to measure elemental abundances.

**Figure 3: EIS plasma composition map.**

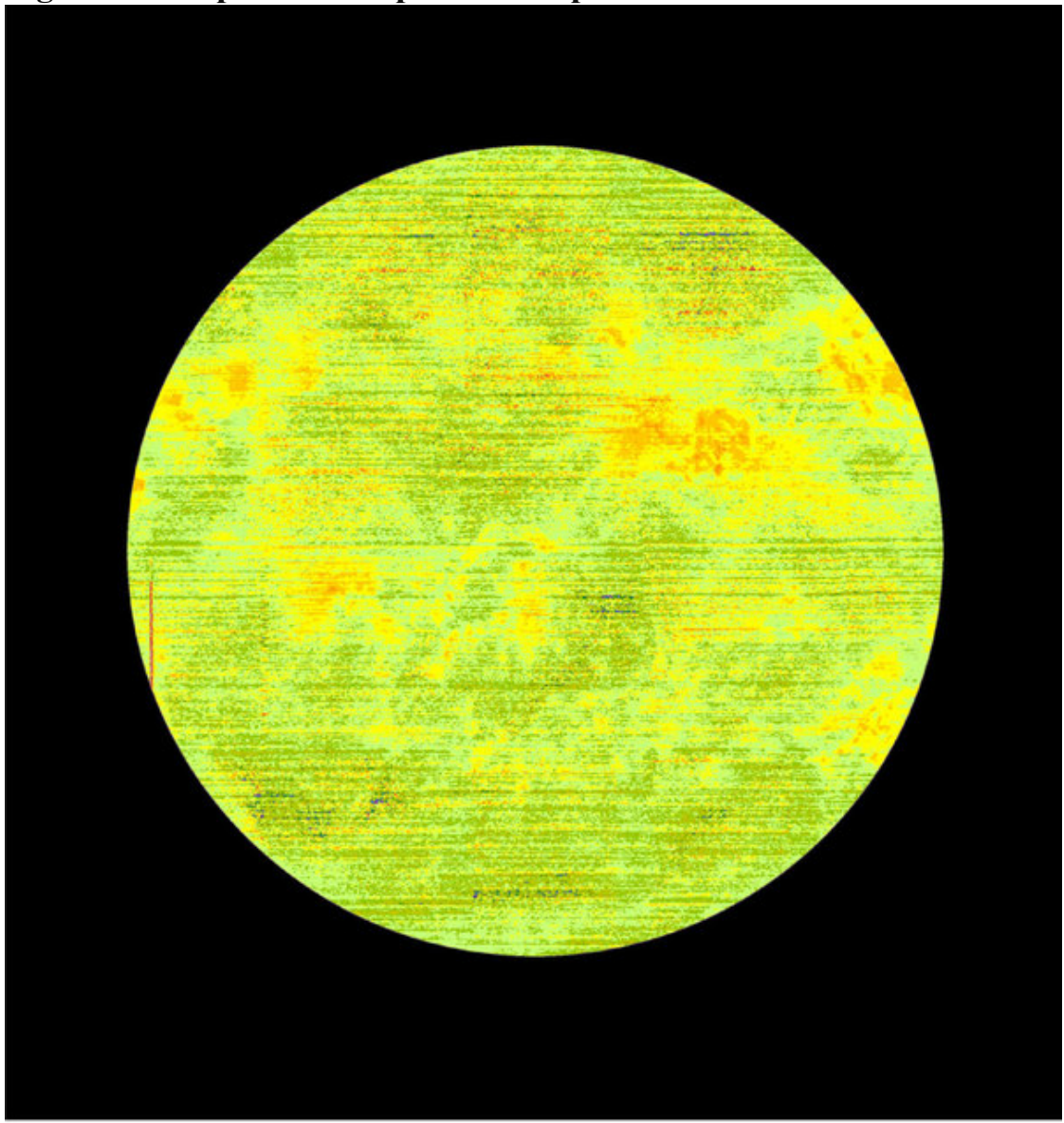

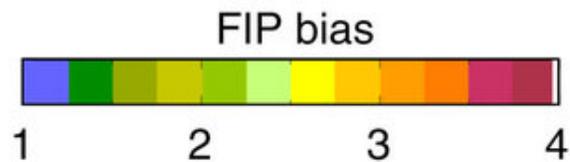

Display version of the full-sun plasma composition map created from the ratio of the Si X 258.37 Å and S X 264.22 Å spectral lines. Darker areas correspond to regions with photospheric abundances. Lighter areas correspond to regions with enhanced (coronal) abundances. To reduce noise, we treated the map using a Fast Fourier Transform filtered by a Hanning mask, and excluded bad pixels and regions outside the solar limb. All of the analysis was performed on the untreated data *(60)*.

**Figure 4: EIS Doppler velocity map of the corona at 2 MK.**

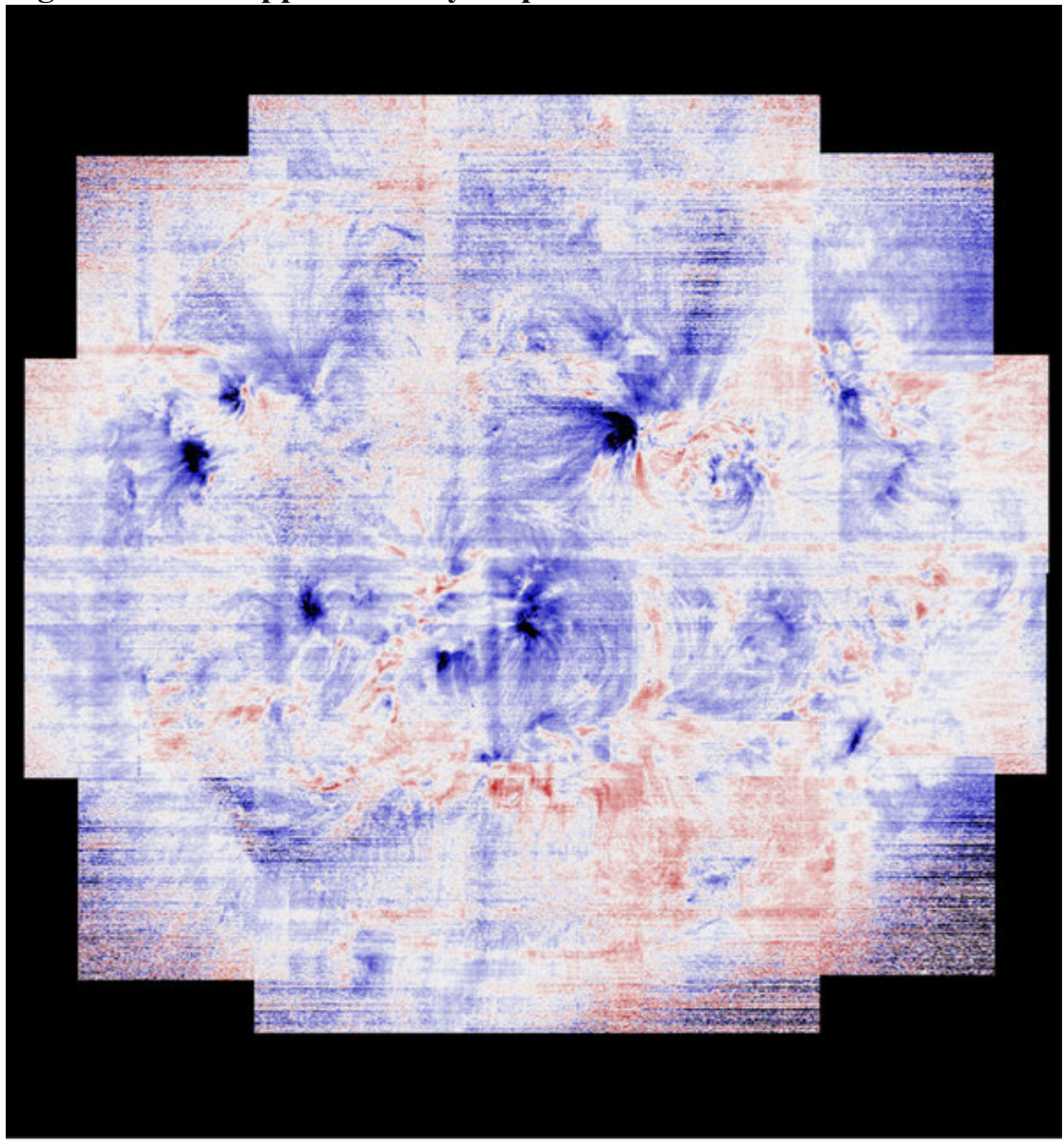

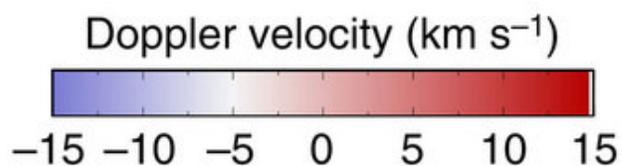

Full-Sun coronal Doppler velocity map derived from single Gaussian fits to the Fe XIII 202.044 Å spectral line. Blue areas highlight plasma that is flowing towards the observer. Red areas highlight plasma that is flowing away from the observer. Vertical artefacts result from the thermal orbital variation of the spectra and should be ignored. The image is scaled to within ±15 km s$^{-1}$. We discuss the details of the velocity derivation method in the main text.

**Figure 5: EIS images of the solar corona with magnetic field lines overlaid.**

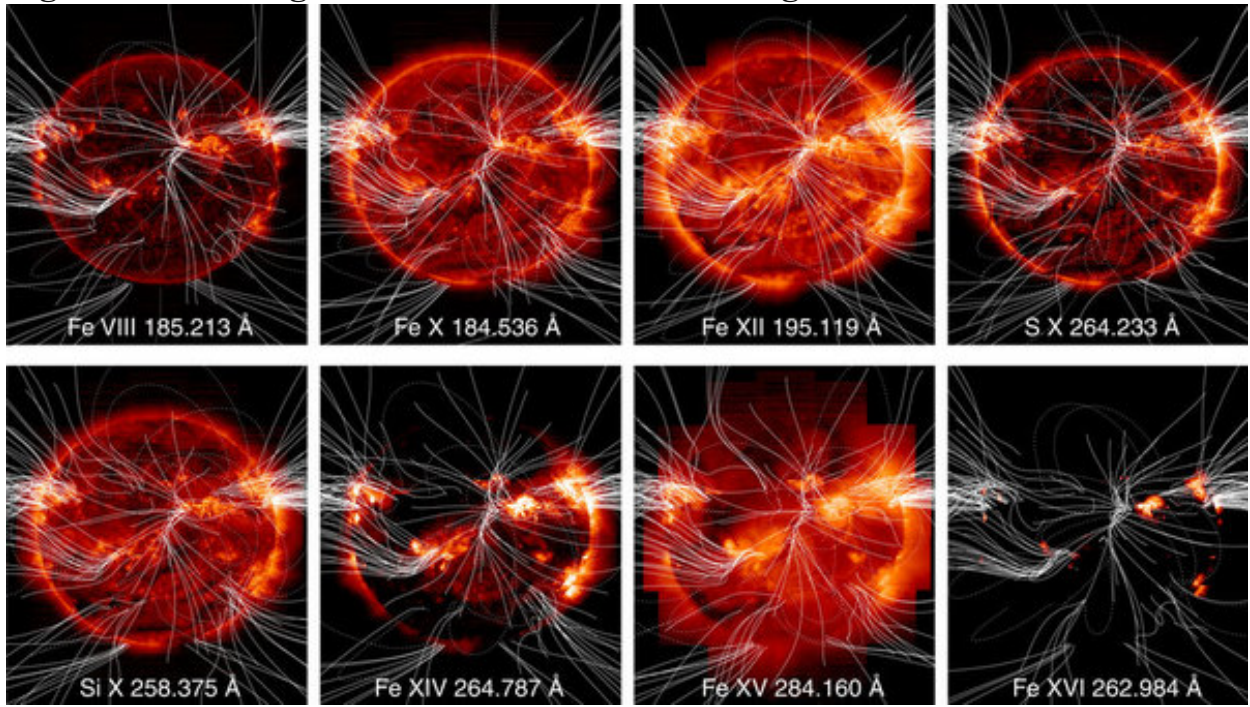

We have overlaid magnetic field lines from our PFSS calculation on the full-Sun intensity images in Fig. 1 that cover a broad range of temperatures from 0.45 to 2.8 MK. The solid field lines are open and the dotted field lines are closed. Only a small subset of the total number of field lines we computed are shown. The extrapolations appear different because the subset was chosen randomly for each intensity image. We have drawn relatively more open-field lines for emphasis.

**Figure 6: EIS Doppler velocity map of the solar corona at 2 MK with magnetic field lines overlaid.**

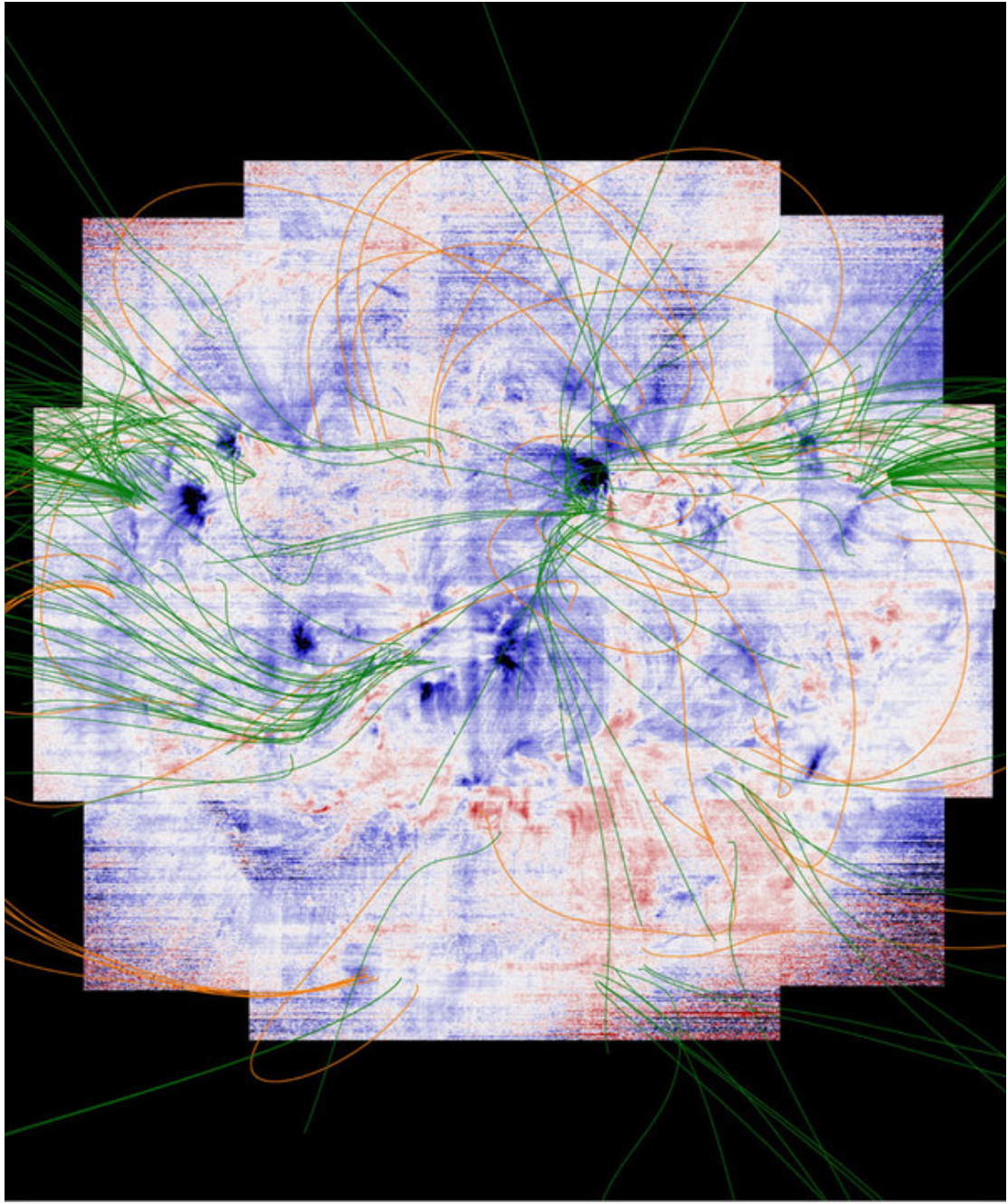

Overlay of magnetic field lines from our PFSS calculation on the Fe XIII 202.044 Å Doppler velocity map in Fig. 4. This time, the green field lines are open and the orange field lines are closed. Again, only a small subset of the total number of field lines we computed are shown: 287 in this case. As with Fig. 5, the field lines are selected randomly and we have drawn relatively more open-field lines for emphasis.

**Figure 7: Relationship between total mass flux and magnetic field line-starting latitude.**

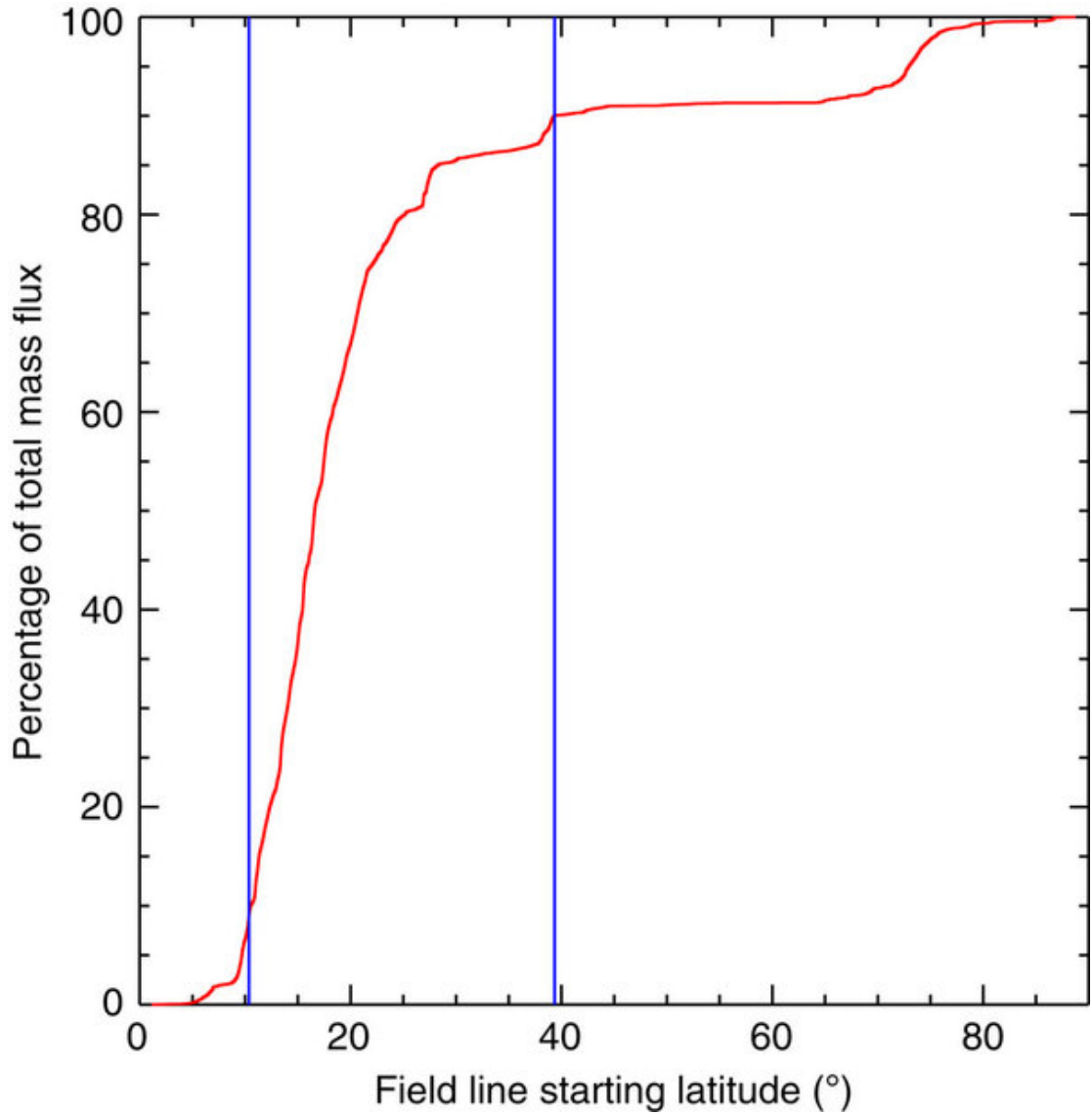

Percentage of total mass flux as a function of magnetic field line-starting latitude (red line). Most of the mass flux (90%) comes from field lines that originate from below 40° latitude (marked by the vertical blue line on the right hand side). The vertical line on the left hand side indicates that most of this flux comes from field lines that originate above a starting latitude of ~11°.

**Figure 8: Slow solar wind source map.**

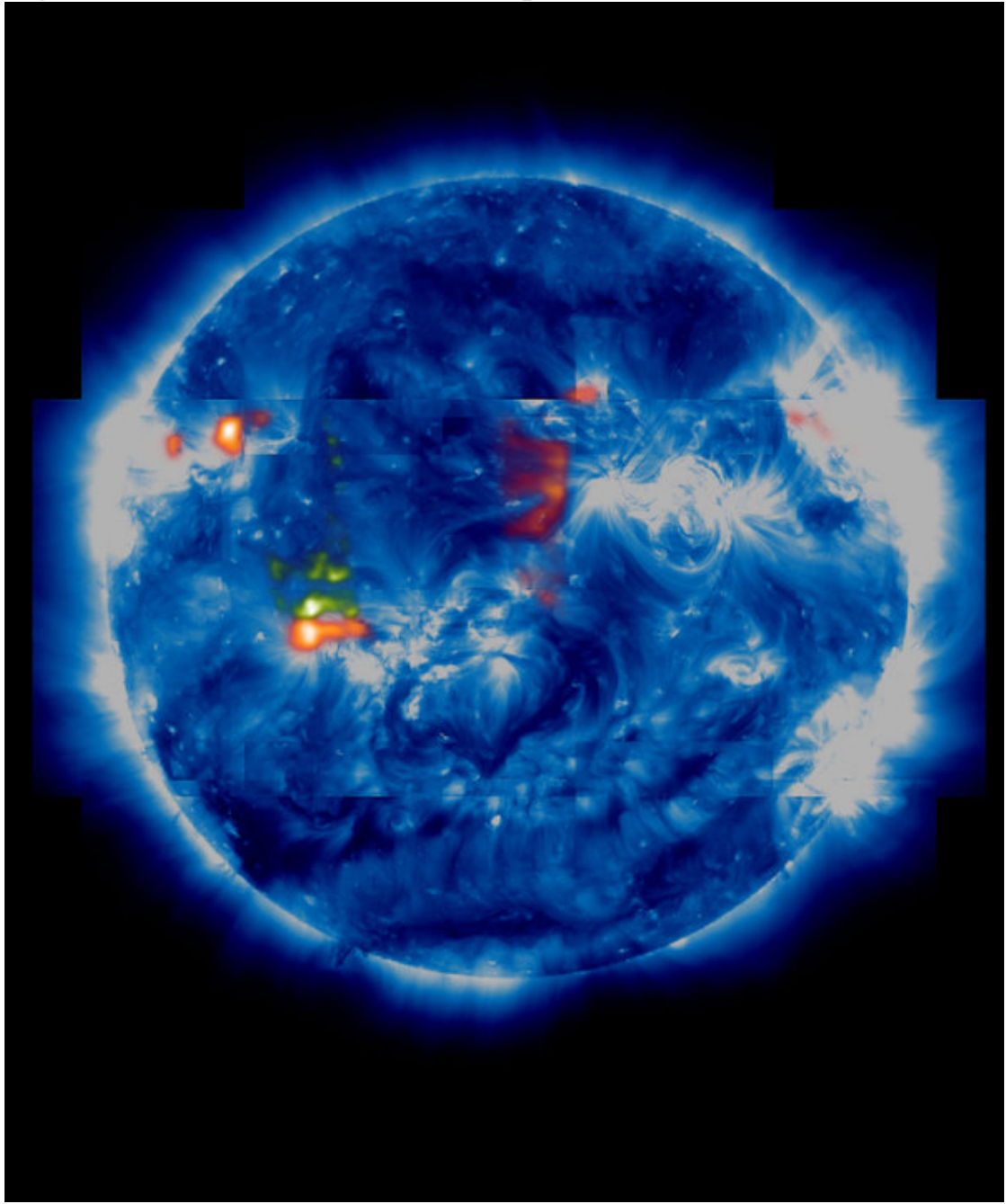

The sources are overlaid on an AIA 193 Å composite intensity image (blue), which we used to correctly coalign and place the EIS raster data in the mosaic. It shows all regions where coronal plasma is outflowing on open-field lines that reach close to the ecliptic plane. These are smoothed with a Gaussian filter to emphasize areas where there is a larger concentration of sources (red). The map is then filtered to identify weaker concentrations, and these are merged on to the image in green. The AIA images have been treated with an unsharp mask to bring out the details.

**Figure 9: Relationship between total mass flux and velocity threshold.**

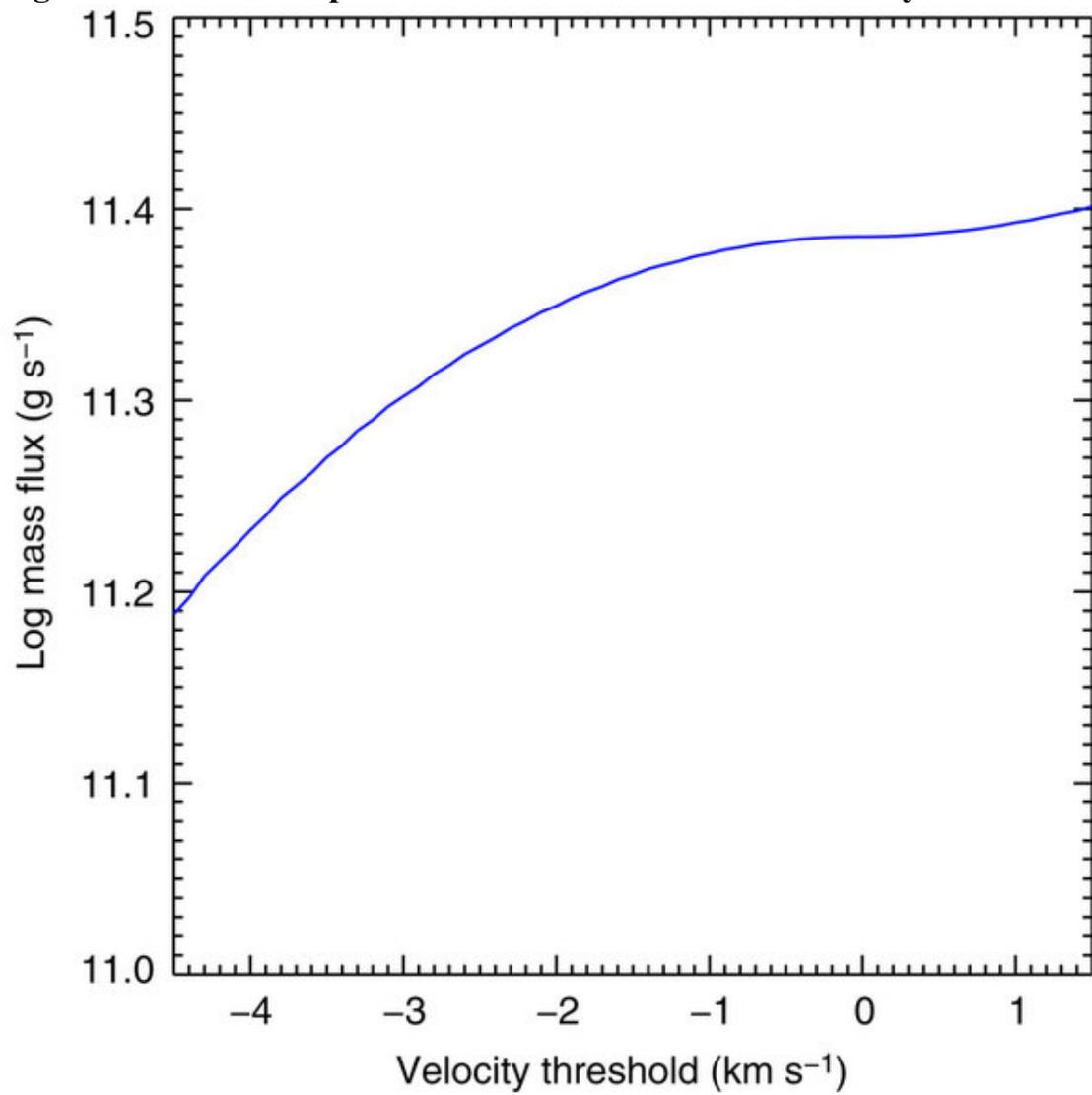

Logarithm of the total mass flux as a function of the chosen velocity threshold.

**Figure 10: ACE measurements of the solar wind near Earth.**

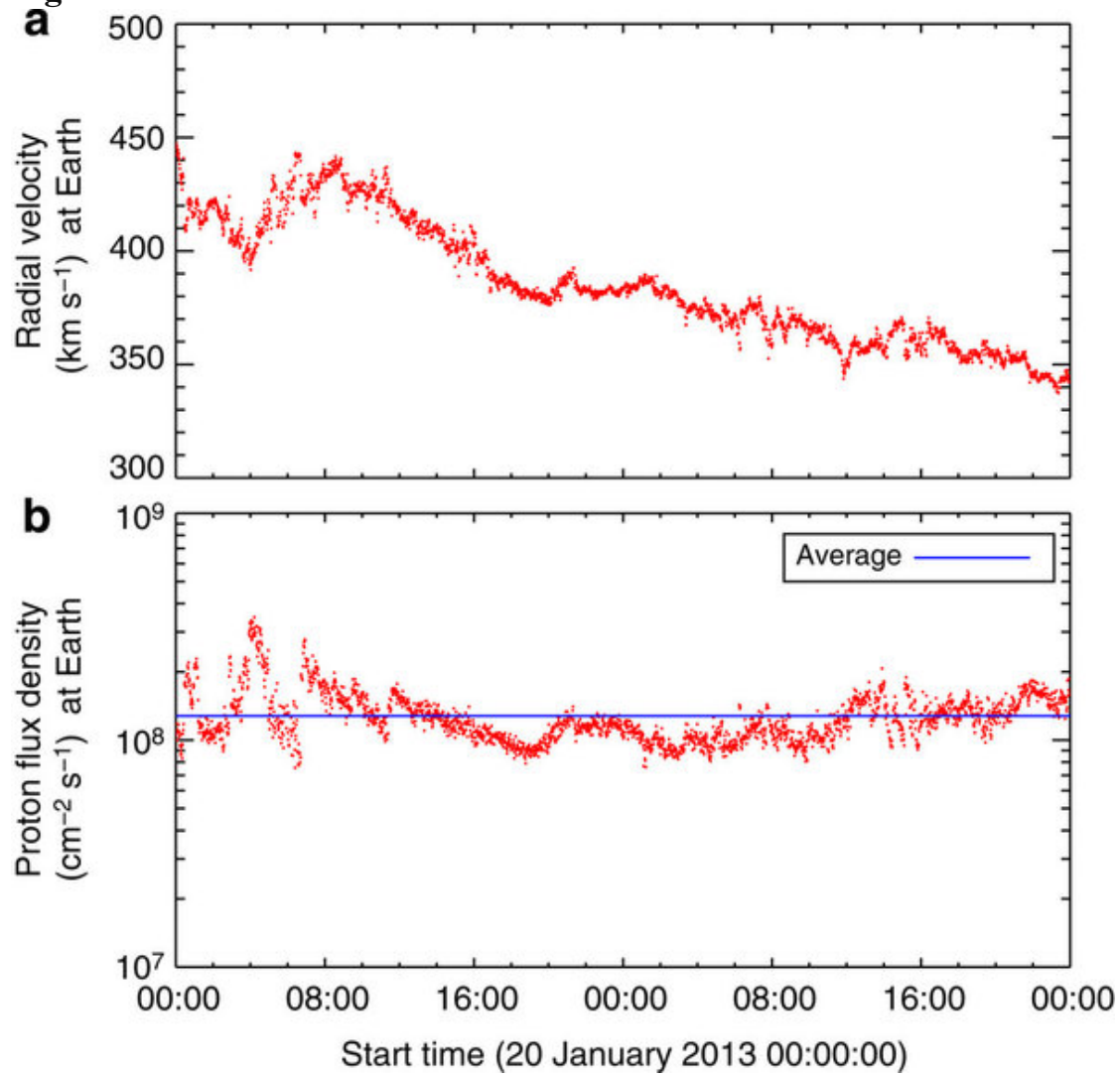

Near-Earth *in situ* radial velocity profile for 20–22 January measured by ACE/SWEPAM (**a**). Proton flux density for the same period (**b**). The average proton flux density is shown as the horizontal blue line.

# Supplementary information

## Methods

**Data reduction**

We treated the EIS data for cosmic rays, dusty, hot and warm pixels and dark current using the standard routine eis_prep, which is available in SolarSoftware. We corrected other instrumental effects such as grating tilt, orbital spectral drift and the spatial offsets between detectors using a neural network model *(44)* and refined the coalignment between different wavelengths using the routine eis_ccd_offset *(59)*. We then calibrated the data to physical units (erg cm$^{-2}$ s$^{-1}$ sr$^{-1}$).

We fit all the spectral lines in the EIS-calibrated scans using single Gaussian functions, except for a few that are blended (the Fe XI and XII lines and Fe XIII 203.826 Å) and so required multiple Gaussian fits to cleanly separate them. We then extracted the intensities for each spectral line at every coaligned pixel.

As discussed in the main text, we improved the placement of the EIS scans within the mosaic using the coaligned full-Sun AIA 193 Å images taken closest in time to the EIS scans. We downloaded these from the Virtual Solar Observatory. They are level 1 data sets and have been corrected for flat fields, cosmic rays and bad pixels and have been converted to DN per pixel s$^{-1}$.

We downloaded the SWEPAM data from the ACE science center at http://www.srl.caltech.edu/ACE/ASC/level2/lvl2DATA_SWEPAM.html. The data are calibrated level 2 and have been verified by the ACE team.

**Test of the FIP bias measurement method**

The simplest method for computing the FIP bias would be to take the S X 264.22 Å to Si X 258.37 Å intensity ratio. These lines are formed at similar temperatures and are very close in wavelength so that calibration issues are minimized. This approach would not, however, account for the temperature and density sensitivity of the line ratio. A more rigorous method is to use density sensitive line ratios to infer the electron density and to use a series of emission lines to compute a temperature distribution. The ratio of the S X 264.22 Å intensity computed from the Differential Emission Measure (DEM) to the observed intensity would yield the FIP bias.

Our analysis method combines these two ideas. We rely primarily on the DEM but introduce a simple scaling using the Si X 258.37 Å line to account for any residual calibration issues because most of the lines used to compute the DEM are on the SW detector, whereas Si and S are on the long-wavelength detector. Using a generative model, we have tested whether the FIP bias determined by our method is sensitive to cross-detector calibration problems, a significant difference in fractionation level between Fe and Si, and uncertainties in the atomic data.

We construct a Gaussian DEM distribution with the peak temperature ($T$) randomly assigned from the range, log $T$=5.8–6.4, and the DEM width ($w$) randomly assigned from the range, log $w$=4.5–5.9. The peak EM is calculated from a randomly assigned density ($n$) in the range log $n$=8.5–9.5. This Gaussian DEM is then used to calculate the intensities of all the spectral lines used in our analysis, and they are then randomly perturbed within the calibration uncertainty. At this point, we can use the generated intensities to calculate the DEM and FIP bias using our analysis method as outlined in the main text.

We performed three sensitivity tests. First, we calculated the FIP bias for 100 simulations, and then, to mimic a significant cross-detector calibration problem, we reduced the SW intensities by a factor of 2 and re-computed the FIP bias factors. We show the results in the left hand panel of Supplementary Fig. 1. The FIP bias factors calculated from the calibration-error model remain within 10% of the original values.

Second, we calculated the FIP bias for 100 simulations, and then, to mimic the effects of a significant difference in fractionation level between Fe and Si, we reduced the Fe abundance by a factor of 2 and re-computed the FIP bias factors. We show the results in the centre panel of Supplementary Fig. 1. The FIP bias factors calculated from the fractionation-difference model remain within 5% of the original values.

Third, to mimic the effects of atomic data uncertainties, we checked the dispersion in FIP bias measurements from 100 simulations that results from increasing the line intensity errors. We show the results of this final test in the right hand panel of Supplementary Fig. 1. Here we found that the FIP bias remains within 30% of the original value until the intensity errors become as large as 40–50%.

In summary, the method we use to determine the FIP bias is robust even if the calibration is significantly in error, Fe and Si are fractionated significantly

differently and/or the uncertainties in the atomic data are as large as 40–50%. PDF files

**Supplementary Figure 1: Test of the FIP bias measurement method**

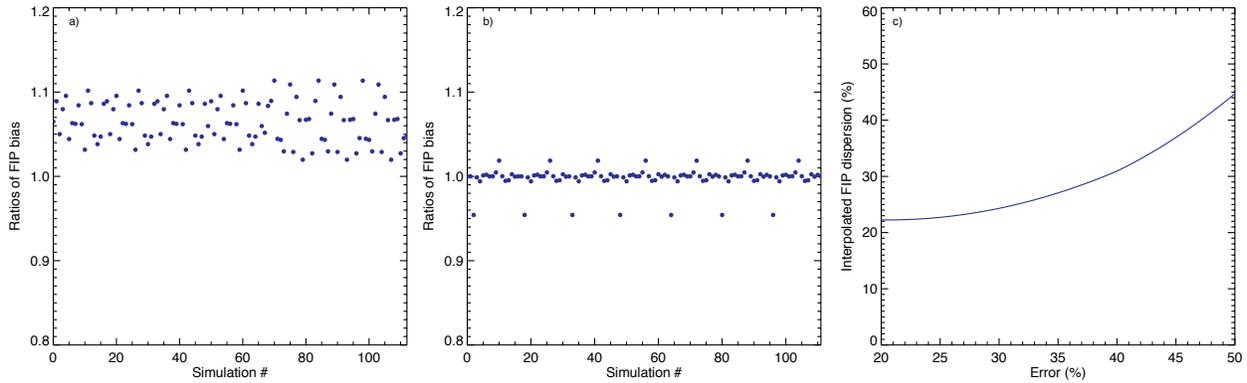

Exploration of the sensitivity of our FIP bias measurements using a generative model. Left panel: ratio of FIP bias factors calculated from 100 simulations with and without reducing the SW intensities by a factor of 2. Center panel: ratio of FIP bias factors calculated from 100 simulations with and without reducing the Fe abundance by a factor of 2. Right panel: dispersion in FIP bias factors from 100 simulations as a percentage of the average FIP bias factor, plotted as a function of the intensity error.